\documentclass[acmsmall]{acmart}
\usepackage{longtable}
\usepackage{booktabs}
\usepackage{geometry}
\usepackage{algorithm}
\usepackage{algpseudocode}
\usepackage{multirow}
\usepackage{pgfplots}
\pgfplotsset{compat=1.18}  % 或您需要的版本
\usepackage{listings}
\usepackage{xcolor}
\AtBeginDocument{%
  }

\setcopyright{acmlicensed}
\copyrightyear{2025}
\acmYear{2025}
\acmDOI{XXXXXXX.XXXXXXX}

\acmJournal{tosem}
\acmVolume{37}
\acmNumber{6}
\acmArticle{111}
\acmMonth{8}

\begin{document}


\title{Hallucination Detection for LLM-based Text-to-SQL Generation via Two-Stage Metamorphic Testing}

\author{Bo Yang}
\email{yangbo@bjfu.edu.cn}
\affiliation{%
  \institution{Beijing Forestry University}
  \city{Haidian}
  \state{Beijing}
  \country{China}
}

\author{Yifen Xia}
\affiliation{%
  \institution{Beijing Forestry University}
  \city{Haidian}
  \state{Beijing}
  \country{China}
}

\author{Weisong Sun}
\authornote{Corresponding author.}
\email{weisong.sun@ntu.edu.sg}
\affiliation{%
 \institution{Nanyang Technological University}
 \country{Singapore}
}

\author{Yang Liu}
\email{yangliu@ntu.edu.sg}
\affiliation{%
 \institution{Nanyang Technological University}
 \country{Singapore}
}

\begin{abstract}
In Text-to-SQL generation, large language models (LLMs) have shown strong generalization and adaptability. However, LLMs sometimes generate hallucinations, i.e.,unrealistic or illogical content, which leads to incorrect SQL queries and negatively impacts downstream applications. Detecting these hallucinations is particularly challenging. Existing Text-to-SQL error detection methods, which are tailored for traditional deep learning models, face significant limitations when applied to LLMs. This is primarily due to the scarcity of ground-truth data. To address this challenge, we propose SQLHD, a novel hallucination detection method based on metamorphic testing (MT) that does not require standard answers. SQLHD splits the detection task into two sequentiial stages: schema-linking hallucination detection via eight structure-aware Metamorphic Relations (MRs) that perturb comparative words, entities, sentence structure or database schema, and logical-synthesis hallucination detection via nine logic-aware MRs that mutate prefix words, extremum expressions, comparison ranges or the entire database. In each stage the LLM is invoked separately to generate schema mappings or SQL artefacts; the follow-up outputs are cross-checked against their source counterparts through the corresponding MRs, and any violation is flagged as a hallucination without requiring ground-truth SQL. The experimental results demonstrate our method's superior performance in terms of the F1-score, which ranges from 69.36\% to 82.76\%. Additionally, SQLHD demonstrates superior performance over LLM Self-Evaluation methods, effectively identifying hallucinations in Text-to-SQL tasks.
\end{abstract}



% \ccsdesc[500]{Software and its engineering~Empirical software validation}



\received{20 February 2025}
\received[revised]{12 March 2025}
\received[accepted]{5 June 2025}
\maketitle
\section{Introduction}
Text-to-SQL generation~\cite{lee2024mcs} has made significant progress in recent years, mainly due to the application of LLMs~\cite{chang2024survey, hadi2023survey, zan2022large}, which have brought new opportunities and challenges to this field. However, the application of LLMs comes with some challenges, the most prominent of which is the phenomenon of hallucinations~\cite{xie2025opensearch, perkovic2024hallucinations, yao2023llm, gekhman2024does, orgad2024LLMs}. Hallucinations in LLMs refer to situations where the content generated by LLMs does not correspond to the real situation, or lacks logical and factual basis. This phenomenon may manifest itself in the form of generating wrong information, fabricating facts that do not exist, or giving conclusions that do not conform to common sense~\cite{huang2025survey, shi2025survey, liu2025survey}. In Text-to-SQL generation, hallucinations directly yield flawed SQL that retrieves incorrect or irrelevant data, seriously endangering downstream applications. For example, in the financial or medical field, incorrect query results can lead to financial loss or medical malpractice. Therefore, how to effectively identify and reduce hallucinations in LLMs is an important research direction in the field of Text-to-SQL generation~\cite{qu2024before, kothyari2023crush4sql}.

% The phenomenon of Text-to-SQL hallucinations based on LLMs is a significant problem~\cite{qu2024before, kothyari2023crush4sql}. Specifically, hallucinations refer to situations where the SQL queries generated by the model do not match the actual database structure or user intent. This can manifest as referencing a table or field that doesn’t exist, generating incorrect logical constraints, or generating SQL queries that are not relevant to the user’s problem. Qu et al.~\cite{qu2024before} divided hallucinations into two categories: schema-based hallucinations and logic-based hallucinations. Schema-based hallucinations refer to the generation of SQL queries that reference tables or fields that do not exist in the database. Logic-based hallucinations, on the other hand, refer to the generation of SQL queries that are logically inconsistent with the user’s query intent.

According to Qu et al.~\cite{qu2024before}, current LLMs for Text-to-SQL exhibit three hallucination classes: schema-based, logic-based and content-based.
\textbf{Schema-based} hallucinations invoke nonexistent tables or columns; detecting them is hard because the model must exactly align question phrases to the true metadata without any labeled schema alignment.
\textbf{Logic-based} hallucinations produce SQL whose clauses contradict the user intent (e.g., reversing ``higher than'' to ⟨ ); the challenge is to spot subtle semantic inversion without executing the query.
\textbf{Content-based} hallucinations filter on values absent from the database (e.g., WHERE pet='rabbit' when no rabbit exists); verification requires inferring the complete value set, which is rarely available at inference time.

At present, the error detection methods for Text-to-SQL tasks using traditional deep learning (DL) methods. DL-based methods train specialized classifiers to recognise mismatches between the question and the candidate SQL by exploiting curated negative samples or execution feedback. These techniques mainly include rule-based detection methods~\cite{wang2023mac}, structured feature learning methods based on graph neural networks~\cite{chen2023error}, and multi-granularity error identification methods~\cite{xu2025boosting}. Rule-based detection methods detect specific types of errors by designing rules. For example, MAC-SQL~\cite{wang2023mac} identifies errors by detecting that SQL execution results are empty or return NULL values. Structured feature learning methods based on graph neural networks learn the structured features of natural language problems and SQL queries through graph neural networks (GNNs),Multi-granularity error identification methods break down SQL errors into finer categories and then detect them separately. 

% Rule-based detection methods detect specific types of errors by designing rules. For example, MAC-SQL~\cite{wang2023mac} identifies errors by detecting that SQL execution results are empty or return NULL values. This type of approach has the advantage of being simple and efficient, but its coverage is limited because it can only detect pre-defined error types. Structured feature learning methods based on graph neural networks learn the structured features of natural language problems and SQL queries through graph neural networks (GNNs), thereby improving the performance and generalization ability of error detection. For example, the error detection model proposed by Chen et al.~\cite{chen2023error} use CodeBERT as the underlying encoder combined with a graph attention network (GAT) to detect errors. Multi-granularity error identification methods break down SQL errors into finer categories and then detect them separately. For example, Xu et al.~\cite{xu2025boosting} proposed a multi-granularity error identification method, which divides SQL errors into system errors, structure errors, and value errors, and designs corresponding detection modules for each category. This method detects syntax errors through an SQL executor and value errors with the help of interaction with the database.

However, the error detection methods for Text-to-SQL tasks that employ traditional deep learning approaches are not readily applicable to hallucination detection in LLMs' Text-to-SQL processes. In addition, a significant challenge in detecting hallucinations in LLMs-based Text-to-SQL tasks is the frequent absence of standard answers (GROUND SQL), which complicates the verification of the correctness of the generated SQL queries. While some studies have proposed using LLMs to detect hallucinations, these methods have limitations in accurately identifying SQL errors because LLMs are not specifically designed to verify the consistency between natural language and SQL queries.

To address this issue, this paper proposes a novel hallucination detection method for Text-to-SQL tasks in LLMs based on MT, termed SQLHD, which operates without requiring a standard answer. SQLHD decomposes the Text-to-SQL task into two distinct phases: schema linking and logical synthesis. 
In the schema-linking phase the LLM API is invoked once to generate a schema-mapping artifact that lists the concrete table–column pairs aligned to the natural-language question; eight structure-aware MRs perturb comparative words, entities, sentence structure, or the database schema, and any mismatch between the original and follow-up mappings is flagged as a schema-level hallucination.
In the logical-synthesis phase the same LLM is invoked again, receiving both the original question and the verified artifact as augmented input, to produce the final SQL query; nine logic-aware MRs of the expected relational behaviour between the source and mutated queries is flagged as a logic-level hallucination.
Because the artifact generated in the first phase is fed forward verbatim, hallucinated tables or columns that escape the first detector are propagated directly into the SQL clauses of the second phase, immediately corrupting downstream synthesis; the two-stage cross-check therefore ensures that undetected hallucinations in the earlier phase are still caught when their logical consequences surface in the SQL.

% Each phase separately invokes the LLMs API to generate different “artifacts,” and the output from the previous phase, along with the original input, is provided to the LLMs as input for the next phase. Consequently, hallucination in the earlier phase directly impact the correctness of subsequent tasks. 

% The contribution of the paper is three-fold.

%  A fine-grained two-stage hallucination detection method is proposed, which decomposes the Text-to-SQL task into two distinct stages: schema linking and logical synthesis. Hallucination detection is performed at each stage to effectively identify and address hallucination problems specific to each phase of the Text-to-SQL task in LLMs.

% Seventeen MRs are introduced to address potential logical and grammatical issues in each stage. These MRs enable the verification of system correctness without relying on explicit standard answers by defining the logical relationship between input and output. This approach effectively characterizes the error patterns specific to different stages of the Text-to-SQL task and provides a robust means for hallucination detection.
	
% Extensive experiments have been conducted, and the results demonstrate that the proposed method effectively detects hallucinations in Text-to-SQL tasks performed by LLMs. It significantly improves the detection rate of hallucinations in both the schema linking and logical synthesis stages, thereby enhancing the overall accuracy of SQL generation.

The major contributions of this work is three-fold:
\begin{itemize}

\item \textbf{A fine-grained two-stage hallucination detection method is proposed}, which decomposes the Text-to-SQL task into two distinct stages: schema linking and logical synthesis. Hallucination detection is performed at each stage to effectively identify and address hallucination problems specific to each phase of the Text-to-SQL task in LLMs.

\item \textbf{We enhanced the precision of error identification in spreadsheets.} SEDMR’s ability to verify MRs allows testers to infer whether the system under test is functioning correctly, regardless of the correctness of the specific implementation. This approach has the potential to reduce the high error rate associated with spreadsheets, which is crucial given their widespread use in various applications.

\end{itemize}

The remainder of this paper is organized as follows. Section~\ref{sec:related work} briefly presents the related work. SQLHD is presented in Section~\ref{sec:approach}. Section~\ref{sec:experiment} present the empirical study including its design. We provide  the experiments results in Section~\ref{sec:experimentresults}. Then, we give the discussions and lessons learned in Section~\ref{sec:discussion}. The paper is concluded in Section~\ref{sec:conclusion}.

\section{Related Work}
\label{sec:related work}
\subsection{LLMs for Text-to-SQL Tasks}
% LLMs have garnered significant attention from both academia and industry for their robust performance and potential in Text-to-SQL tasks~\cite{liu2025survey, shi2025survey, hong2024next, yang2024synthesizing, zhang2024sqlfuse}. The development of these models is increasingly demonstrating substantial value and notable achievements across various domains. This technology primarily leverages the models' capabilities in understanding and generating natural language, as well as their grasp of SQL syntax and semantics, to construct an efficient and accurate system for converting natural language queries into SQL queries.
LLMs have demonstrated strong cross-domain generalization in tasks that require both natural-language understanding and structured output generation, and this capability extends to Text-to-SQL conversion where they map user questions to executable SQL by exploiting their in-context grasp of relational semantics and query syntax~\cite{liu2025survey,shi2025survey,hong2024next,yang2024synthesizing,zhang2024sqlfuse}.

The application of LLMs in Text-to-SQL tasks has evolved from simple prompt engineering to more complex fine-tuning methods. For instance, early research such as that by Dong et al.~\cite{dong2023c3} proposed C3, which utilized zero-shot prompting to harness the pre-trained knowledge of LLMs for Text-to-SQL tasks. As research has progressed, more methods have begun to employ few-shot learning and chain-of-thought techniques to enhance model performance. For example, Pourreza et al.~\cite{pourreza2023din} introduced DIN-SQL, which improves the accuracy of SQL generation by decomposing complex queries and utilizing chain-of-thought reasoning. Additionally, some studies have started to explore how fine-tuning can further boost the performance of LLMs in Text-to-SQL tasks. For instance, Pourrez et al. ~\cite{pourreza2024sql} proposed SQL-GEN, which fine-tunes LLMs on specific task datasets to enhance their ability to generate domain-specific SQL queries. Currently, an increasing number of studies are focusing on combining prompt engineering and fine-tuning methods to achieve better performance. For example, Zhang et al.~\cite{zhang2024finsql} introduced FinSQL, which fine-tunes LLMs on financial domain datasets and integrates prompt engineering techniques to efficiently generate SQL queries in the financial sector. These Text-to-SQL systems based on LLMs provide technical support for the intelligent transformation of database querying. 

Xie et al.~\cite{xie2024decomposition} introduce the workflow prompting paradigm to enhance the performance of LLMs in text-to-SQL tasks. This approach leverages structured prompts to guide the model through a series of reasoning steps, thereby improving the accuracy and reliability of SQL query generation.
Li et al.~\cite{li2025omnisql} study the problem of prompt designing in the text-to-SQL task and attempt to improve the LLMs' reasoning ability when generating SQL queries. Their work focuses on crafting effective prompts that can better align the model's output with the desired SQL structure.
Liu et al.~\cite{liu2024solid} propose SolidSQL, a pre-processing pipeline designed to enhance the robustness of LLM-based text-to-SQL systems. SolidSQL incorporates data cleaning, schema alignment, and query normalization techniques to improve the overall stability and accuracy of the model's SQL output.
Caferoğlu et al.~\cite{caferouglu2024sql} introduce E-SQL, a novel pipeline specifically designed to address the challenges of schema linking and predicate augmentation in text-to-SQL tasks. E-SQL employs direct schema linking and candidate predicate augmentation to enhance the model's ability to generate accurate SQL queries.
Cao et al.~\cite{cao2024rsl} introduce RSL-SQL, a framework for generating text-to-SQL using LLMs. RSL-SQL focuses on improving the model's understanding of relational schemas and query semantics through a combination of reinforcement learning and self-supervised learning techniques.
Shen et al.~\cite{shen2025magesql} propose MageSQL, a new framework for text-to-SQL based on in-context learning over LLMs. MageSQL utilizes context-aware learning to adapt the model to specific query patterns and database schemas, thereby enhancing its ability to generate contextually relevant SQL queries.
Li et al.~\cite{li2024codes} introduce CodeS, a series of pre-trained language models with parameters ranging from 1B to 15B, specifically designed for the text-to-SQL task. CodeS models are fine-tuned on large-scale text-to-SQL datasets to improve their performance in generating accurate and efficient SQL queries.

\subsection{MT}

MT is a software testing methodology that does not directly verify whether the output of a software is correct. Instead, it examines whether the software's behavior remains consistent after applying a series of predefined MRs. A MR refers to the expected relationship between inputs and outputs when the target program is executed multiple times. It is the core concept of MT, used to generate follow-up test cases and validate the correctness of the program by comparing the results of multiple executions. For example, for a program that calculates the sine value, the MR can be $\sin$($\pi$ - $x$) = $\sin$($x$). Even without knowing the exact output for a given input, the correctness of the program can be verified through this relationship.

As an effective technique for addressing the ``Oracle Problem'' in software testing, MT has achieved significant development and wide application. Chen et al.~\cite{chan1998application} introduced the basic methodology and theoretical framework of MT, proposing the use of MRs to generate follow-up test cases to verify software correctness. Barr et al.~\cite{barr2014oracle} provided a comprehensive review of the ``Oracle Problem,'' emphasizing the significance of MT. Additionally, Chen et al.~\cite{chen2020metamorphic} applied MT in the field of numerical analysis, demonstrating its effectiveness in specific domains. In addition, Chen et al.~\cite{chen2003choice} proposed a test case generation method based on the Category-Choice Framework, providing a theoretical basis for the subsequent generation of MRs.

Regarding the generation and optimization of MRs, Dong et al.~\cite{dong2008case} studied the combination method of MRs, generating new ones by combining existing relations, offering an approach for test case diversification. Kanewala et al.~\cite{kanewala2013using} introduced machine learning techniques into the prediction of MRs, improving the accuracy of predictions by constructing control flow graphs and introducing graph kernel techniques. Xie et al.~\cite{xie2020mettle} developed a metamorphic testing approach to assess and validate unsupervised machine learning systems. Ayerdi et al.~\cite{ayerdi2021generating} proposed a MR auto-generation method based on genetic programming, providing a new pathway for automated generation. Zhang et al.~\cite{zhang2023automated} used ChatGPT to generate MRs, showcasing the potential of LLMs in this domain. Deng et al.~\cite{deng2021bmt} proposed a MT framework based on Behavior-Driven Development (BDD), specifically for the testing of autonomous driving systems. This framework integrates MT with BDD, offering a novel solution for testing in specific domains.  In 2016, Chen et al.~\cite{chen2018metamorphic} reviewed the challenges and opportunities of MT, summarizing its successful cases in various domains. Liu et al.~\cite{liu2013effectively} conducted an in-depth study on the effectiveness of MT in mitigating the ``Oracle Problem.'' Li et al.~\cite{li2024metamorphic} reviewed the latest research progress in the generation of MRs and proposed future research directions.

\section{Methodology}
\label{sec:approach}
\subsection{The framework of SQLHD}
\begin{figure*}[!htb]
	\includegraphics[width=0.99\linewidth]{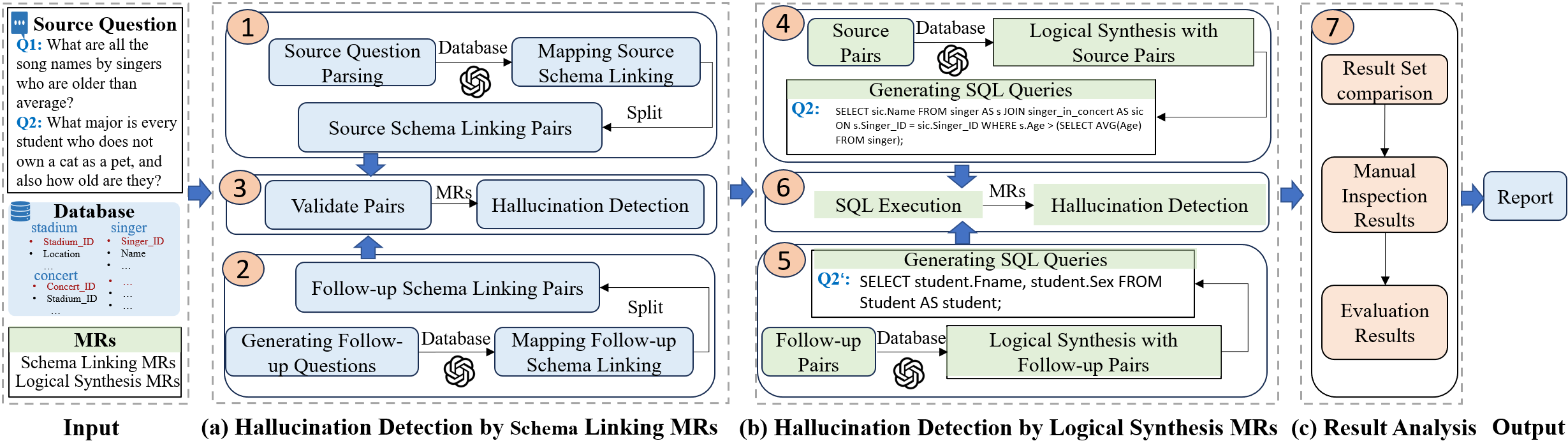}
	\caption{The SQLHD framework }
    \label{fig:SQLHD}
\end{figure*}

SQLHD utilizes a two serial phases methodology to identify and rectify hallucinations in LLM-based Text-to-SQL generation, as depicted in Figure~\ref{fig:SQLHD}. SQLHD is structured into discrete stages, each designed to pinpoint and counteract hallucinations at various junctures within the Text-to-SQL conversion process. The process is initiated with three fundamental inputs: Source Questions, a Database schema, and MRs. The Source Questions are natural language queries intended to be transformed into SQL queries by the system. The Database schema delineates the data's organizational structure, including tables and fields, which is crucial for accurately mapping the questions to their corresponding database fields. MRs further facilitate this process by establishing connections between the question elements and the database schema, and by directing the logical construction of SQL queries from the questions. These inputs lay the groundwork for the framework's procedural steps, encompassing parsing, schema linking, validation, and hallucination detection, all geared towards ensuring the precision of the SQL queries produced. Subsequently, hallucination detection is executed through schema linking MRs. Should the schema linking MRs fail to detect hallucinations, the process escalates to hallucination detection via logical synthesis MRs. Ultimately, the detection outcomes are analyzed to compile a comprehensive detection report.

\subsection{Schema Linking MRs}
Schema linking grounds a natural language question in the target database by aligning phrases to concrete tables, columns and relationships.  This step is critical because any misalignment, such as selecting a nonexistent table or overlooking the correct join path, directly propagates into an erroneous clause and invalidates the final query.  The main difficulties are lexical ambiguity, where the same phrase may match multiple schema items, structural sparsity, since most columns are never mentioned in training examples, and hallucination sensitivity, because large language models freely generate unseen table or column names when confidence is low. 

To facilitate hallucination detection in schema linking, we have proposed eight MRs, categorized into four distinct groups based on the extent of metamorphosis relations: Comparative Word Metamorphosis (CWM), Entity Metamorphosis (ENM), Sentence Metamorphosis (SEM), and Database Schema Metamorphosis (DSM). CWM includes Synonym Replacement of Comparative Words (SROCW) and Antonym Replacement of Comparative Words (AROCW). ENM includes Entity Synonymous Replacement (ESR) and Entity Non-Synonymous Replacement (ENSR). SEM includes Remove Conditional (RC), Ordinary Simplification (OS), and Sentence Complication (SC). DSM consists of Database Schema Replacement (DSR). This classification across multiple dimensions, including semantic understanding, sentence structure, and database schema adaptability, thereby facilitating a thorough detection and analysis of hallucinations in LLMs. SROCW refers to the process of generating a Follow-up Question by replacing the comparative word in the Question with a synonym. SROCW is defined as follows:

\[
  \text{SROCW:}
  \begin{cases}
  \text{SQ}=\langle sr_1,s_2,\ldots,sr_t,\ldots,sr_{n-1},sr_n\rangle\\
  \text{FUQ}=\langle sr_1,sr_2,\ldots,sr'_t,\ldots,sr_{n-1},sr_n\rangle\\
  \end{cases}
  \]
where $SQ$ is a resource question. $FUQ$ is the follow-up question. These definitions are consistent across all the MRs described. $sr\_t$ is a comparative word and $sr'\_t$ is a synonym of $sr\_t$.

AROCW defines a particular type of MR where the comparative words in the source question are replaced with their antonyms. Comparative words are typically used to express comparisons in quantity or degree, such as older than, lower than, greater than, etc. AROCW is defined as follows:
\[
  \text{AROCW:}
  \begin{cases}
  \text{SQ}=\langle ar_1,ar_2,\ldots,ar_t,\ldots,ar_{n-1},ar_n\rangle\\
  \text{FUQ}=\langle ar_1,ar_2,\ldots,ar'_t,\ldots,ar_{n-1},ar_n\rangle\\
  \end{cases}
  \]
where \( ar_t \) is a comparative word and \( ar'_t \) is an antonym of \( ar_t \).

% By utilizing these two metamorphosis relationships, we select the comparative words in the source question (Source Question) and replace them with synonyms and antonyms to generate follow-up questions (Follow-up Question). The distinguishing feature of the question pairs generated at this metamorphosis level is that their sentence structures are highly similar, with only individual comparative words undergoing metamorphosis. Importantly, the change in comparative words does not affect the generation of Schema Linking. Therefore, the Schema Linking generated by the Source Question and the Follow-up Question should remain consistent. If the Schema Linking generated by the Source Question and the Follow-up Question is inconsistent, it indicates the presence of hallucinations in the Schema Linking generated by LLMs.

ESR involves generating a follow-up question by replacing entities in the source question with their synonyms. ESR is defined as follows:
\[
  \text{ESR:}
  \begin{cases}
  \text{SQ}=\langle es_1,es_2,\ldots,es_t,\ldots,es_{n-1},es_n\rangle\\
  \text{FUQ}=\langle es_1,es_2,\ldots,es'_t,\ldots,es_{n-1},es_n\rangle\\
  \end{cases}
  \]
where \( es_t \) is an entity and \( es'_t \) is a synonym of \( es_t \).

ENSR is defined as follows:
\[
  \text{ENSR:}
  \begin{cases}
  \text{SQ}=\langle en_1,en_2,\ldots,en_t,\ldots,en_{n-1},en_n\rangle\\
  \text{FUQ}=\langle en_1,en_2,\ldots,en'_t,\ldots,en_{n-1},en_n\rangle\\
  \end{cases}
  \]
where \( en_t \) is an entity and \( en'_t \) is a non-synonym of \( en_t \).

To ESR and ENSR, If the Schema Linking generated for the source question and the follow-up question are inconsistent, it indicates the presence of hallucinations in the Schema Linking generated by LLMs.

RC,OS and SC extend beyond specific words or phrases, indicating that the Source Question and Follow-up Question may differ not only in meaning but also in sentence structure. RC is defined as follows:
\[
  \text{RC:}
  \begin{cases}
  \text{SQ}=\langle rc_1,rc_2,\ldots,rc_t,\ldots,rc_{n-1},rc_n\rangle\\
  \text{FUQ}=\langle rc_1,rc_2,\ldots,rc_{n-1},rc_n\rangle\\
  \end{cases}
  \]
where \( rc_t \) is a conditional statement.

OS is defined as follows:
\[
  \text{OS:}
  \begin{cases}
  \text{SQ}=\langle os_1,os_2,\ldots,os_t,\ldots,os_{n-1},os_n\rangle\\
  \text{FUQ}=\langle os_1,os_2,\ldots,os_{n-1},os_n\rangle\\
  \end{cases}
  \]
where \( os_t \) is an irrelevant component.

SC is defined as follows:
\[
  \text{SC:}
  \begin{cases}
  \text{SQ}=\langle sc_1,sc_2,\ldots,sc_t,\ldots,sc_{n-1},sc_n\rangle\\
  \text{FUQ}=\langle sc_1,sc_2,\ldots,sc'_t,\ldots,sc_{n-1},sc_n\rangle\\
  \end{cases}
  \]
where \( sc_t \) is an irrelevant component and \( sc'_t \) is additional content.

% The MR of the RC removes conditional statements from the Source Question, potentially reducing the number of entities in the sentence. This process enhances the accuracy of LLMs’ comprehension of the problem, necessitating that the schema linking generated by the Source Question encompasses the schema linking produced by the Follow-up Question.

% This MR of the OS involves the removal of irrelevant components from the Source Question, ensuring that key entity names are not reduced in the sentence. This process makes the problem more concise, thereby requiring the Schema Linking generated by the Source Question to be identical to the Schema Linking generated by the Follow-up Question. LLMs assist in removing these irrelevant components from the Source Questionand are guided by a small number of examples to delete irrelevant components in the Question and generate a new query statement, resulting in the Follow-up Question.

% Conversely, the MR  of SC adds an irrelevant component to the Source Question, without increasing the number of key entity names in the sentence. This makes the problem more verbose, necessitating a higher level of semantic understanding from LLMs. This metamorphosis relationship also requires the Schema Linking generated by the Source Question to be the same as the Schema Linking generated by the Follow-up Question. LLMs assist in adding irrelevant components to Source Questions, and are guided by a small number of examples to add irrelevant components to the Question and generate new query statements, resulting in the Follow-up Question.

DSR denotes changes to the database schema, serving as the basis for formulating follow-up questions. DSR is defined as follows:
\[
  \text{DSR:}
  \begin{cases}
  \text{SQ}=\langle dsr_1,dsr_2,\ldots,dsr_t,\ldots,dsr_{n-1},dsr_n\rangle\\
  \text{FUQ}=\langle dsr'_1,dsr'_2,\ldots,dsr'_t,\ldots,dsr'_{n-1},dsr'_n\rangle\\
  \end{cases}
  \]
where \( dsr_t \) is a table in the database schema and \( dsr'_t \) is a new table resulting from the transformation of \( dsr_t \). At least one table has been modified.

\subsection{Hallucination Detection by Schema Linking MRs}
Source Question Parsing is a pivotal initial step that involves meticulously analyzing the input query to uncover its underlying structural elements and semantic content. 
% During this phase, the parsing process identifies critical information such as entities, attributes, conditions, and the intentions conveyed within the source question. 
% This thorough extraction of details is essential for accurately aligning the question with the database schema and for crafting follow-up questions that accurately reflect the original intent without introducing any logical fallacies or inaccuracies.
In the Generating Follow\-up Questions, a comprehensive set of MRs is employed to produce a series of follow\-up questions. These MRs include CWM, ENM,SEM and DSM. The Validate Pairs phase involves a rigorous examination of each pair of source and follow\-up questions to ensure they preserve logical consistency and semantic relevance. Mapping Source Schema Linking maps the components of the parsed source question onto the corresponding elements within the database schema. Following validation and mapping, the Split phase separates the validated pairs into source and follow\-up questions. Finally, the system applies the defined MRs to meticulously scrutinize both the source and follow\-up questions for any inconsistencies or deviations. Algorithm 1 illustrates the details of hallucination detection based on schema linking MRs.
% It takes four inputs: a source question (SQ), its associated database schema (SDS), a follow-up question (FUQ) derived through MRs, and Schema Linking MRs (SLMRs). 

% \begin{algorithm}[!htb]
% \caption{Hallucination Detection in Schema Linking}
% \label{alg:hallucination_detection_schema_linking}
% \begin{algorithmic}
% \Require{Input:} $SQ$, $SDS$, $FUQ$, $SLMRs$\;
% \Require{Output:} $HDs$, $CM$\;

% \State $SSL \gets GenerateSourceSL(SQ, SDS)$\;
% \State $FSL \gets GenerateFollowUpSL(FUQ, FDS)$\;
% \State $HDs \gets 0$\;
% \State $CM \gets InitializeConfusionMatrix()$\;

% \For{each $MR$ in $SLMRs$}
%     \If{$MR \in \{SROCW, AROCW, ESR, RC, OS, SC\}$}
%         \If{$SSL(MR) \neq FSL(MR)$}
%             \State $HDs$ \gets $HDs$ + $1$\;\\
%             \State Update $CM$\;
%         \Else
%             \State Skip this MR\;
%         \EndIf
%     \Else
%         \If{$MR \in \{ENSR, DSR\}$}
%             \If{$SSL(MR) == FSL(MR)$}
%                 \State $HDs$ \gets $HDs$ + $1$\;\\
%                 \State Update $CM$\;
%             \Else
%                 \State Skip this MR\;
%             \EndIf
%         \EndIf
%     \EndFor
    
% \Return $HDs$, $CM$\;
% \end{algorithmic}
% \end{algorithm}
\begin{algorithm}[!htb]
\caption{Hallucination Detection in Schema Linking}
\label{alg:hallucination_detection_schema_linking}
\begin{algorithmic}[1]
\Require $SQ$, $SDS$, $FUQ$, $SLMRs$
\Ensure $HDs$, $CM$

\State $SSL \gets \text{GenerateSourceSL}(SQ, SDS)$
\State $FSL \gets \text{GenerateFollowUpSL}(FUQ, SDS)$
\State $HDs \gets 0$
\State $CM \gets \text{InitializeConfusionMatrix}()$

\For{each $MR$ in $SLMRs$}
    \If{$MR \in \{SROCW, AROCW, ESR, RC, OS, SC\}$}
        \If{$SSL(MR) \neq FSL(MR)$}
            \State $HDs \gets HDs + 1$
            \State Update $CM$
        \Else
            \State Skip this MR
        \EndIf
    \Else
        \If{$MR \in \{ENSR, DSR\}$}
            \If{$SSL(MR) == FSL(MR)$}
                \State $HDs \gets HDs + 1$
                \State Update $CM$
            \Else
                \State Skip this MR
            \EndIf
        \EndIf
    \EndIf
\EndFor

\State \Return $HDs$, $CM$
\end{algorithmic}
\end{algorithm}
\subsection{Logical Synthesis MRs}
Logical synthesis translates the grounded schema into an executable SQL query by inserting the correct join paths, selection filters, grouping clauses and aggregate expressions. This stage is error-prone because the model must reconcile relational algebra with the nuances of the question, including implicit ordering, quantifier scope and extremum constraints. Typical failures are erroneous join topology, misuse of aggregate functions and reversal of comparative direction, all of which remain syntactically valid and thus evade simple syntax checks. To surface these logical faults we propose four categories of MRs that mutate prefix words, extremum expressions, comparison ranges or the entire database; any deviation between the source query result and the follow-up result reveals a hallucination without relying on ground-truth SQL.

To facilitate hallucination detection in logical synthesis, we proposed four categories of MRs: Prefix Word Metamorphose (PWM), Extrema Metamorphose (EXM), Comparative Range Metamorphose (CRM), and Database Metamorphose (DAM). PWM includes Prefix Insertion (PI), Prefix Removal (PR), and Prefix Substitution (PS). EXM includes Synonym Replacement of Extrema (SROE) and Antonym Replacement of Extrema (AROE). CRM includes Comparison Range Unchanged (CRU), Comparison Range Expand (CRE), and Comparative Words Reduce (CWR). DAM consists of Database Replacement (DR). This classification spans multiple dimensions such as semantic precision, logical coherence, extremum handling, and adaptability to database schema changes, which are critical for ensuring the reliability of SQL query generation. 

PI refers to adding a prefix word to the sentence, like Tell me or Show me. These additions typically don't alter the sentence's meaning but can influence how the model interprets and generates the query. PI is defined as follows:
\[
  \text{PI:}
  \begin{cases}
  \text{SQ}=\langle pi_1,pi_2,\ldots,pi_t,\ldots,pi_{n-1},pi_n\rangle\\
  \text{FUQ}=\langle \text{prefix}, pi_1,pi_2,\ldots,pi_t,\ldots,pi_{n-1},pi_n\rangle\\
  \end{cases}
  \]
where \( pi_t \) is a component of the source question and ``prefix'' is the inserted prefix word.

PR refers to removing a prefix word, making the sentence more concise but potentially affecting the model's understanding of its completeness. PR is defined as follows:

\[
  \text{PR:}
  \begin{cases}
  \text{SQ}=\langle pr_1,pr_2,\ldots,pr_t,\ldots,pr_{n-1},pr_n\rangle\\
  \text{FUQ}=\langle pr_1,pr_2,\ldots,pr_{n-1},pr_n\rangle\\
  \end{cases}
  \]
where \( pr_t \) is a prefix word in the source question.

PS refers to replacing one prefix word with another, which might change the tone or style but not the core semantics. PS is defined as follows:

\[
  \text{PS:}
  \begin{cases}
  \text{SQ}=\langle ps_1,ps_2,\ldots,ps_t,\ldots,ps_{n-1},ps_n\rangle\\
  \text{FUQ}=\langle ps_1,ps_2,\ldots,ps'_t,\ldots,ps_{n-1},ps_n\rangle\\
  \end{cases}
  \]
where \( ps_t \) is a prefix word in the source question and \( ps'_t \) is the substituted prefix word.

SROE refers to replacing an extreme word with a synonym, such as changing ``highest'' to ``maximum.'' This doesn't change the meaning but tests if the model recognizes these synonyms. SROE is defined as follows:

\[
  \text{SROE:}
  \begin{cases}
  \text{SQ}=\langle sroe_1,sroe_2,\ldots,sroe_t,\ldots,sroe_{n-1},sroe_n\rangle\\
  \text{FUQ}=\langle sroe_1,sroe_2,\ldots,sroe'_t,\ldots,sroe_{n-1},sroe_n\rangle\\
  \end{cases}
  \]
where \( sroe_t \) is an extremum word (e.g., ``highest'') and \( sroe'_t \) is a synonym of \( sroe_t \).

AROE refers to replacing an extreme word with an antonym, like changing ``highest'' to ``lowest.'' This significantly changes the meaning and tests the model's ability to handle logical reversals. AROE is defined as follows:

\[
  \text{AROE:}
  \begin{cases}
  \text{SQ}=\langle aroe_1,aroe_2,\ldots,aroe_t,\ldots,aroe_{n-1},aroe_n\rangle\\
  \text{FUQ}=\langle aroe_1,aroe_2,\ldots,aroe'_t,\ldots,aroe_{n-1},aroe_n\rangle\\
  \end{cases}
  \]
where \( aroe_t \) is an extremum word (e.g., ``highest'') and \( aroe'_t \) is an antonym of \( aroe_t \).

CRU refers to substituting a comparative word with a synonym, like replacing ``greater than'' with ``exceeds,'' without altering the meaning. CRU is defined as follows:

Comparison Range Unchanged (CRU)
\[
  \text{CRU:}
  \begin{cases}
  \text{SQ}=\langle cru_1,cru_2,\ldots,cru_t,\ldots,cru_{n-1},cru_n\rangle\\
  \text{FUQ}=\langle cru_1,cru_2,\ldots,cru'_t,\ldots,cru_{n-1},cru_n\rangle\\
  \end{cases}
  \]
where \( cru_t \) is a comparative word (e.g., ``older than'') and \( cru'_t \) is a synonym of \( cru_t \).

CRE refers to broadening the comparative word's range, such as changing ``greater than'' to ``not less than,'' which increases the result set size. CRE is defined as follows:

\[
  \text{CRE:}
  \begin{cases}
  \text{SQ}=\langle cre_1,cre_2,\ldots,cre_t,\ldots,cre_{n-1},cre_n\rangle\\
  \text{FUQ}=\langle cre_1,cre_2,\ldots,cre'_t,\ldots,cre_{n-1},cre_n\rangle\\
  \end{cases}
  \]
where \( cre_t \) is a comparative word (e.g., ``older than'') and \( cre'_t \) is a word that expands the comparison range.

CWR refers to narrowing the comparative word's range, like changing ``greater than'' to ``exceeds,'' which decreases the result set size. CWR is defined as follows:

\[
  \text{CWR:}
  \begin{cases}
  \text{SQ}=\langle cwr_1,cwr_2,\ldots,cwr_t,\ldots,cwr_{n-1},cwr_n\rangle\\
  \text{FUQ}=\langle cwr_1,cwr_2,\ldots,cwr'_t,\ldots,cwr_{n-1},cwr_n\rangle\\
  \end{cases}
  \]
where \( cwr_t \) is a comparative word (e.g., ``more than'') and \( cwr'_t \) is a word that reduces the comparison range.

DR refers to swapping one database schema for another, completely changing the model's context and testing its ability to adapt to a new schema. DR is defined as follows:

\[
  \text{DR:}
  \begin{cases}
  \text{SQ}=\langle dr_1,dr_2,\ldots,dr_t,\ldots,dr_{n-1},dr_n\rangle\\
  \text{FUQ}=\langle dr'_1,dr'_2,\ldots,dr'_t,\ldots,dr'_{n-1},dr'_n\rangle\\
  \end{cases}
  \]
where \( dr_t \) is a table or schema element in the source database and \( dr'_t \) is the corresponding element in the replaced database.

\subsection{Hallucination Detection via Logical Synthesis MRs}
The Hallucination Detection phase initiates with the Logical Synthesis of the Source Question, where an initial SQL query is generated based on the parsed source question. This query is meticulously designed to extract data from the database that aligns with the user's original intent. Following the initial logical synthesis, the follow-up Schema Linking phase ensures that the transformed follow\-up question maintains correct associations with the database fields. 
The SQL Execution phase involves executing the synthesized SQL queries against the database to retrieve result sets. This step is pivotal as it provides the actual output that can be analyzed for accuracy and consistency, offering a tangible basis for evaluating the performance of the generated SQL queries against the user's intent.

For example, the SQL query for the Follow-up Question ''Show me what major is every student who does not own a cat as a pet, and also how old are they?'' is:
% \begin{lstlisting}[language=SQL]
% SELECT s.Major, s.Age FROM Student AS s LEFT JOIN Has_Pet AS hp ON s.StuID = hp.StuID AND NOT EXISTS ( SELECT 1 FROM Pets p WHERE p.PetType = 'cat' AND p.PetID = hp.PetID );
% \end{lstlisting}

\begin{lstlisting}[language=SQL]
SELECT s.Major, s.Age
FROM Student AS s
LEFT JOIN Has_Pet AS hp
    ON s.StuID = hp.StuID
    AND NOT EXISTS (
        SELECT 1
        FROM Pets p
        WHERE p.PetType = 'cat'
          AND p.PetID = hp.PetID
    );
\end{lstlisting}
This query results in a set size of 35. Another example is the query for the Follow-up Question ''What major is every student who does not own a cat as a pet, and also how old are they?'' which is:
% \begin{lstlisting}[language=SQL]
% SELECT s.name AS Major, s.age AS Age FROM students AS s WHERE s.pet_type != 'cat';
% \end{lstlisting}

\begin{lstlisting}[language=SQL]
SELECT s.name AS Major, s.age AS Age
FROM students AS s
WHERE s.pet_type != 'cat';
\end{lstlisting}
However, this query results in an error or no result, indicated by a set size of -1.

Post-execution, the Hallucination Detection phase employs specific MRs related to logical synthesis to compare the results of the source and follow-up questions.

\subsection{Generating Hallucination Detection Report}
In this Step, SQLHD meticulously compares the result sets from the executed SQL queries to detect any discrepancies that might signify hallucinations in data retrieval. Upon identifying discrepancies, a thorough manual inspection is conducted to understand their nature and determine whether they stem from hallucinations in the SQL queries or are expected due to the different questions asked. In addition, SQLHD evaluates the results to assess the accuracy and reliability of the generated and executed SQL queries. 
\section{Experiments}
\label{sec:experiment}
\subsection{Research Questions} 
\textbf{RQ1:} What is the effectiveness of SQLHD?
The calculation of hallucinations based on overall MRs aims to validate the effectiveness of the method proposed and provide a baseline comparison for evaluating the hallucination detection capability of individual MRs. To this end, we select four LLMs: GLM-4, ChatGPT (gpt-3.5-turbo), DeepSeek-V3, and Qwen2.5-32B, and conduct experiments on the two representative datasets, Spider and BIRD. By comparing and analyzing the experimental data, we can clearly demonstrate the performance of our method in hallucination detection, thereby providing strong empirical support for subsequent research.

\textbf{RQ2:} What is the effectiveness of the schema-based MR hallucination detection method for LLMs Text-to-SQL tasks?

During the schema linking phase, LLMs precisely map the parsing components of natural language queries to the corresponding tables, columns, and relationships in the database schema. This phase is the core of the entire query parsing process because it directly determines whether the system can accurately understand and interpret the query intent in the correct context. Given the complexity of semantic understanding and structural matching involved in schema linking, hallucinations (such as incorrect mappings or misinterpretations of query intent) are highly likely to occur, thereby affecting the accuracy of subsequent SQL generation. Therefore, evaluating the effectiveness of hallucination detection methods based on schema-based MRs is crucial.

\textbf{RQ3:}  What is the effectiveness of the logic-based MR hallucination detection method for LLMs Text-to-SQL tasks?

In the logic synthesis phase, we input the results of schema linking from the previous stage, along with other relevant information, into the large language model to generate the final SQL statement. This phase involves executing precise JOIN operations, applying appropriate SQL clauses, and conducting complex computational reasoning in data science queries. Given the complexity of these tasks and the high demands on the model's reasoning capabilities, hallucinations (i.e., generating incorrect or irrelevant information) are prone to occur. Therefore, evaluating the effectiveness of hallucination detection methods based on logic-based MRs is essential. This not only helps us identify and correct errors but also enhances the accuracy and reliability of the generated SQL statements, better serving the needs of complex queries in practical applications.

\textbf{RQ4:} What is the effectiveness of the hallucination detection method based on individual MRs for LLMs Text-to-SQL tasks?

This study systematically evaluates the hallucination detection performance of individual MRs, aiming to reveal the detection characteristics of different types of MRs through multidimensional empirical analysis. The study focuses on examining the differential performance of various MRs in hallucination detection tasks, combining quantitative metrics with qualitative analysis to not only precisely assess the detection sensitivity and applicability of each type but also to deeply identify the advantages and limitations of their internal detection mechanisms.

% \textbf{RQ5:} What is the effectiveness of the hallucination detection method based on a single MR for LLMs Text-to-SQL tasks?

% Evaluating the hallucination detection performance of a single MR aims to provide a comprehensive understanding of the detection capabilities of each MR. This evaluation not only helps us fully understand the performance of each MR in hallucination detection tasks but also reveals their strengths and weaknesses. Through detailed analysis, we can identify which MRs perform well in specific scenarios and which need further optimization, thereby providing a solid basis for improving model performance. This is of great significance for enhancing the overall detection efficiency and accuracy of MRs and lays a solid foundation for subsequent research and applications.

\subsection{Datasets} 
We conducted experiments on two widely used Text-to-SQL dev datasets: the Spider dataset developed by the Computer Science Department at Yale University~\cite{yu2018spider}, and the BIRD dataset jointly developed by the Alibaba Group Damo Academy and several universities in China~\cite{li2023can}. The specific details of the datasets are shown in Table~\ref{tab:dataset_details}. Leveraging GLM-4 and ChatGPT, we augmented the original Spider and BIRD corpora, yielding four mutated datasets: SGLMD and BGLMD (GLM-4-driven), alongside SGPMD and BGPMD (ChatGPT-driven).

\begin{table*}[]
\centering
\caption{Dataset Details}
\label{tab:dataset_details}
\begin{tabular}{@{}ccccl@{}}
\toprule
Dataset                 & Type     & Tables & SQL pairs & Description                                 \\ \midrule
\multirow{3}{*}{Spider} & training & 140    & 7000      & \multirow{3}{*}{\begin{tabular}[c]{@{}l@{}}The dataset needs the model \\ to generalize to new SQL\\  queries and database schemas.\end{tabular}}                       \\
                        & develop  & 20     & 1034      &                                                                                                                                                                                   \\
                        & test     & 6      & 1659      &                        \\ \midrule
\multirow{3}{*}{BIRD}   & training & 69     & 9428      & \multirow{3}{*}{\begin{tabular}[c]{@{}l@{}}The dataset incorporates dirty \\ data and noise values, and also \\ considers external knowledge.\end{tabular}} \\
                        & develop  & 11     & 1534      &                                                                                                                                                                                   \\
                        & test     & 15     & 1789      &                                                                                                                                                                                   \\ \bottomrule
\end{tabular}
\end{table*}

\subsection{Evaluation Metrics}
For RQ1 and RQ4, we chose the \textsl{Precision}, \textsl{Recall}, \textsl{F1-score}~\cite{Opitz2019MacroFA} and \textsl{Hallucination Detection Number} (\textsl{HDN}) to evaluate the effectiveness of SQLHD. There quantifiable metrics for hallucination detection in Text-to-SQL tasks based on MRs for LLMs.

\begin{enumerate}
    \item \textbf{Hallucination Detection Number (HDN)}\\
    Total number of hallucinations caught by \emph{any} MR in the LLM output.
    \begin{equation}
        \mathrm{HDN}=|\{\text{hallucination instances}\}|
    \end{equation}

    \item \textbf{Hallucination Detection Rate (HDR)}\\
    Ratio of hallucinations caught by a \emph{specific} MR to those caught by its parent MR set.
    \begin{equation}
        \mathrm{HDR}=\frac{\mathrm{HDN}_{\text{MR}_i}}{\mathrm{HDN}_{\text{parent}}},
        \quad \mathrm{HDN}_{\text{parent}}\neq 0
    \end{equation}

    \item \textbf{Accuracy}\\
    Proportion of samples whose final SQL is judged hallucination-free.
    \begin{equation}
        \mathrm{Accuracy}=\frac{\#\,\text{Correct (no-hallucination) predictions}}{\text{Total \# samples}}
    \end{equation}

    \item \textbf{Precision}\\
    Among all samples flagged as positive (no hallucination), how many were truly positive.
    \begin{equation}
        \mathrm{Precision}=\frac{\mathrm{TP}}{\mathrm{TP}+\mathrm{FP}}
    \end{equation}

    \item \textbf{Recall}\\
    Among all ground-truth positive samples, how many were correctly identified.
    \begin{equation}
        \mathrm{Recall}=\frac{\mathrm{TP}}{\mathrm{TP}+\mathrm{FN}}
    \end{equation}

    \item \textbf{F1-Score}\\
    Harmonic mean of Precision and Recall.
    \begin{equation}
        \mathrm{F1-score}=2\cdot\frac{\mathrm{Precision}\cdot\mathrm{Recall}}{\mathrm{Precision}+\mathrm{Recall}}
    \end{equation}
\end{enumerate}

\section{Results and Analyses}
\label{sec:experimentresults}
\subsection{RQ1 -- Analysis of the Effectiveness of the Overall MR-Based Hallucination Detection Method}

The experimental results of the overall MR-based hallucination detection method are shown in Table~\ref{tab:overall_metamorphic_effectiveness}.

\begin{table*}[]
\centering
\caption{The Effectiveness of Overall MRs in Detecting Hallucinations}
\label{tab:overall_metamorphic_effectiveness}
\begin{tabular}{cccccc}
\hline
\textbf{Dataset}       & \textbf{LLMs/Method} & \textbf{Accuracy (\%)} & \textbf{Precision (\%)} & \textbf{Recall (\%)} & \textbf{F1 (\%)} \\ \hline
\multirow{5}{*}{SGLMD} & GLM-4                & 95.55                  & 62.50                   & 94.59                & 75.27            \\
                       & TA-SQL+ GLM-4        & 96.62                  & \textbf{72.41}          & 96.55                & \textbf{82.76}   \\
                       & ChatGPT              & 95.07                  & 71.55                   & 82.18                & 76.50            \\
                       & DeepSeek-V3          & 89.75                  & 55.51                   & 96.18                & 70.39            \\
                       & Qwen2.5-32B          & 94.29                  & 70.72                   & 95.52                & 81.27            \\ \hline
\multirow{5}{*}{SGPMD} & GLM-4                & 91.78                  & 62.03                   & 89.23                & 73.19            \\
                       & TA-SQL+ GLM-4        & 93.81                  & 66.29                   & 95.87                & 78.38            \\
                       & ChatGPT              & 91.97                  & 64.56                   & 93.01                & 76.22            \\
                       & DeepSeek-V3          & 88.82                  & 66.88                   & 94.50                & 78.33            \\
                       & Qwen2.5-32B          & 92.67                  & 70.90                   & 97.74                & 82.19            \\ \hline
\multirow{5}{*}{BGLMD} & GLM-4                & 96.54                  & 54.55                   & 95.24                & 69.36            \\
                       & TA-SQL+ GLM-4        & \textbf{98.31}         & 63.33                   & 90.48                & 74.51            \\
                       & ChatGPT              & 97.20                  & 68.25                   & 96.63                & 80.00            \\
                       & DeepSeek-V3          & 93.29                  & 55.14                   & 94.40                & 69.62            \\
                       & Qwen2.5-32B          & 95.57                  & 64.48                   & 97.52                & 77.63            \\ \hline
\multirow{5}{*}{BGPMD} & GLM-4                & 94.98                  & 55.88                   & 97.94                & 71.16            \\
                       & TA-SQL+ GLM-4        & 95.70                  & 62.35                   & \textbf{98.15}       & 76.26            \\
                       & ChatGPT              & 92.18                  & 59.23                   & 98.27                & 73.91            \\
                       & DeepSeek-V3          & 92.79                  & 60.38                   & 95.73                & 74.06            \\
                       & Qwen2.5-32B          & 93.35                  & 59.23                   & 95.17                & 73.02            \\ \hline
\end{tabular}
\end{table*}

From the comparison results in Table~\ref{tab:overall_metamorphic_effectiveness} we observe that the unified MR pipeline consistently delivers F1-scores between 69\% and 83\% across all four datasets and five large-scale models. The narrow spread indicates that the method is largely insensitive to schema complexity or model family, and therefore offers a stable, plug-and-play safeguard for Text-to-SQL generation.

Second, the best single-model result (TA-SQL + GLM-4 on SGLMD, 82.76\% F1-score) is only 3.4 percentage points above the worst performer (GLM-4 on BGPMD, 69.36\% F1), demonstrating that the seven MRs compensate for each other's blind spots and prevent any one model from dominating the outcome. Such robustness is essential for production environments where the downstream SQL generator may be upgraded frequently.

Finally, the recall of every configuration exceeds 89\%, while precision stays above 54\%, showing that the overall MR set rarely misses a hallucination and keeps false alarms at a manageable level. Taken together, these findings confirm that our metamorphic-relation-based detector provides a reliable, model-agnostic mechanism to improve the accuracy and trustworthiness of large language models when they synthesise SQL queries.

\textbf{In conclusion, expanding MRs through the SQLHD framework effectively enhances the detection of hallucinations in Text-to-SQL generation. More diverse MRs provide broader coverage of schema-linking and logical-synthesis scenarios, increasing the likelihood of identifying subtle hallucinations that might be missed with a smaller relation set.}

\subsection{RQ2 -- Analysis of the Effectiveness of the Schema-Linking MR-Based Hallucination Detection Method}
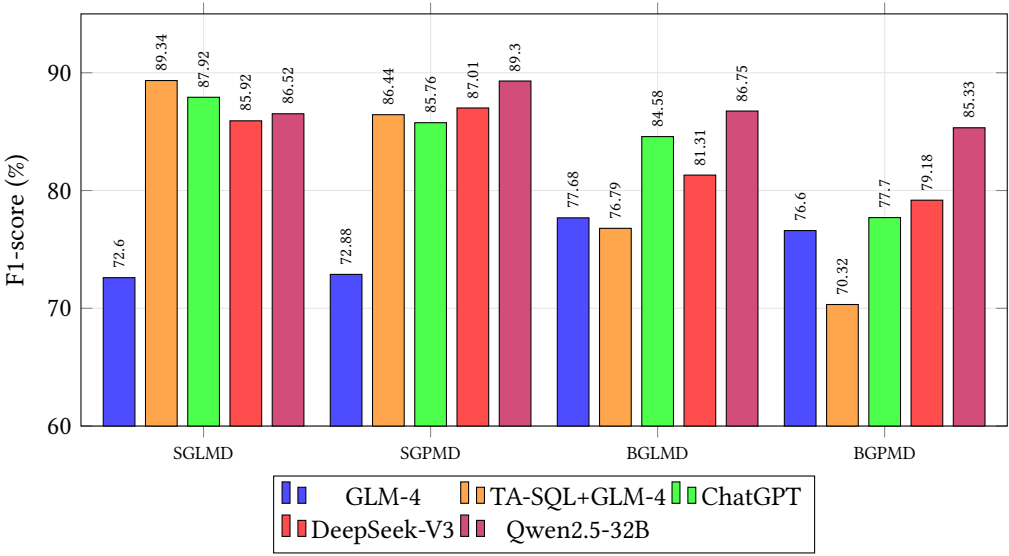
\begin{figure*}[t]
\centering
\begin{tikzpicture}
\begin{axis}[
    width=\linewidth,
    height=7cm,
    ybar=4pt,
    bar width=12pt,
    enlarge x limits=0.18,
    ymin=60, ymax=95,
    ylabel={F1-score (\%)},
    xlabel={Dataset},
    % 一行标签
    symbolic x coords={SGLMD, SGPMD, BGLMD, BGPMD},
    xtick=data,
    x tick label style={font=\scriptsize},
    nodes near coords,
    every node near coord/.append style={font=\tiny, rotate=90, anchor=west},
    legend style={at={(0.5,-0.12)}, anchor=north, legend columns=3},
    grid=major,
    grid style={gray!20}
]

% GLM-4
\addplot[fill=blue!70,draw=black] coordinates {
    (SGLMD,72.60) (SGPMD,72.88) (BGLMD,77.68) (BGPMD,76.60)
};

% TA-SQL+GLM-4
\addplot[fill=orange!70,draw=black] coordinates {
    (SGLMD,89.34) (SGPMD,86.44) (BGLMD,76.79) (BGPMD,70.32)
};

% ChatGPT
\addplot[fill=green!70,draw=black] coordinates {
    (SGLMD,87.92) (SGPMD,85.76) (BGLMD,84.58) (BGPMD,77.70)
};

% DeepSeek-V3
\addplot[fill=red!70,draw=black] coordinates {
    (SGLMD,85.92) (SGPMD,87.01) (BGLMD,81.31) (BGPMD,79.18)
};

% Qwen2.5-32B
\addplot[fill=purple!70,draw=black] coordinates {
    (SGLMD,86.52) (SGPMD,89.30) (BGLMD,86.75) (BGPMD,85.33)
};

\legend{GLM-4, TA-SQL+GLM-4, ChatGPT, DeepSeek-V3, Qwen2.5-32B}
\end{axis}
\end{tikzpicture}
\caption{The Effectiveness of Hallucination Detection Based on Schema-Linking MRs}
\label{tab:schema_linking_metamorphic_effectiveness}
\end{figure*}

The experimental results of RQ2 are shown in Figure~\ref{tab:schema_linking_metamorphic_effectiveness}. It can be observed from Figure~\ref{tab:schema_linking_metamorphic_effectiveness} that the schema-linking phase is crucial in Text-to-SQL tasks as it involves mapping natural language queries to the corresponding database schema components (tables, columns, and relationships). This phase is where many hallucinations can occur due to incorrect mappings or misunderstandings of the query intent. The schema-linking MR-based hallucination detection method leverages this phase to identify inconsistencies and potential hallucinations.
2. Performance Across Datasets
The method shows consistent performance across different datasets (SGLMD, SGPMD, BGLMD, and BGPMD), with F1 scores ranging from 70.32\% to 89.34\%. This indicates that the method is robust and can effectively detect hallucinations in various contexts.

\textbf{\textit{From the comparison results, we can observe that for all four datasets, our schema-linking MR approach achieved higher performance. This suggests that our approach is effective in identifying hallucinations in Text-to-SQL tasks, providing a reliable method to enhance the accuracy and reliability of LLMs in generating SQL queries.}} 

\subsection{RQ3 -- The Effectiveness of Hallucination Detection Based on Logical Synthesis MRs}

\begin{figure*}[t]
\centering
\begin{tikzpicture}
\begin{axis}[
    width=\linewidth,
    height=7cm,
    ybar=4pt,
    bar width=12pt,
    enlarge x limits=0.18,
    ymin=60, ymax=90,
    ylabel={F1 (\%)},
    xlabel={Dataset},
    symbolic x coords={SGLMD,SGPMD,BGLMD,BGPMD},
    xtick=data,
    x tick label style={font=\scriptsize},
    nodes near coords,
    every node near coord/.append style={font=\tiny, rotate=90, anchor=west},
    legend style={at={(0.5,-0.18)}, anchor=north, legend columns=3},
    grid=major,
    grid style={gray!20}
]

% GLM-4
\addplot[fill=blue!70,draw=black] coordinates {
    (SGLMD,76.50) (SGPMD,75.08) (BGLMD,69.36) (BGPMD,71.16)
};

% TA-SQL+GLM-4
\addplot[fill=orange!70,draw=black] coordinates {
    (SGLMD,82.76) (SGPMD,78.38) (BGLMD,76.00) (BGPMD,76.26)
};

% ChatGPT
\addplot[fill=green!70,draw=black] coordinates {
    (SGLMD,78.30) (SGPMD,78.24) (BGLMD,80.00) (BGPMD,73.91)
};

% DeepSeek-V3
\addplot[fill=red!70,draw=black] coordinates {
    (SGLMD,70.79) (SGPMD,79.02) (BGLMD,69.62) (BGPMD,74.06)
};

% Qwen2.5-32B
\addplot[fill=purple!70,draw=black] coordinates {
    (SGLMD,81.79) (SGPMD,82.58) (BGLMD,77.63) (BGPMD,73.02)
};

\legend{
    GLM-4, TA-SQL+GLM-4, ChatGPT,
    DeepSeek-V3, Qwen2.5-32B
}
\end{axis}
\end{tikzpicture}
\caption{The Effectiveness of Hallucination Detection Based on Logical Synthesis MRs}
\label{tab:logical_synthesis_metamorphic_effectiveness}
\end{figure*}
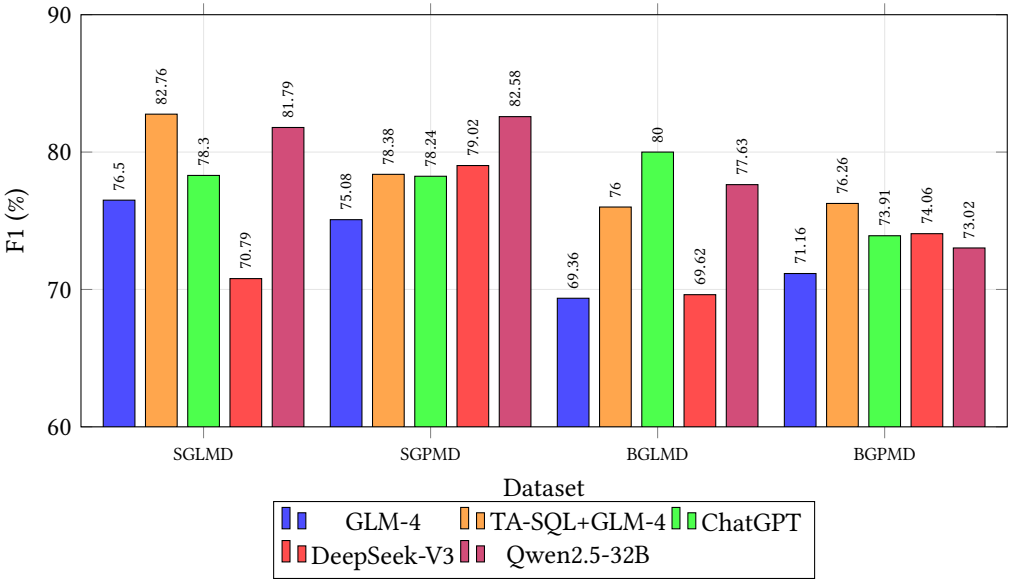

The experimental results of RQ3 are shown in Figure~\ref{tab:logical_synthesis_metamorphic_effectiveness}.
It can be seen from Figure~\ref{tab:logical_synthesis_metamorphic_effectiveness} that the method demonstrates a relatively stable performance across different metamorphic datasets and LLMs, with the F1-score ranging from 69.36\% to 82.76\%. Additionally, the high recall values indicate a low false-negative rate, suggesting that the method is effective in identifying hallucinations.

The logical synthesis phase is a critical step in Text-to-SQL tasks, where the parsed components from the schema-linking phase are used to generate the final SQL query. This phase involves executing precise JOIN operations, applying appropriate SQL clauses, and conducting complex computational reasoning. Due to the complexity of these tasks, hallucinations (i.e., generating incorrect or irrelevant information) are prone to occur. The logical synthesis MR-based hallucination detection method leverages this phase to identify inconsistencies and potential hallucinations by ensuring that the generated SQL queries adhere to the logical constraints defined by the MRs.

\textbf{\textit{From the comparison results, we can observe that for all four datasets, our logical synthesis MR-based approach achieved higher performance. This suggests that our approach is particularly effective in identifying and mitigating hallucinations in Text-to-SQL tasks.}} 

\subsection{RQ4 -- Ablation study about single-class MRs}
\begin{figure}[!htb]
	\includegraphics[width=0.99\linewidth]{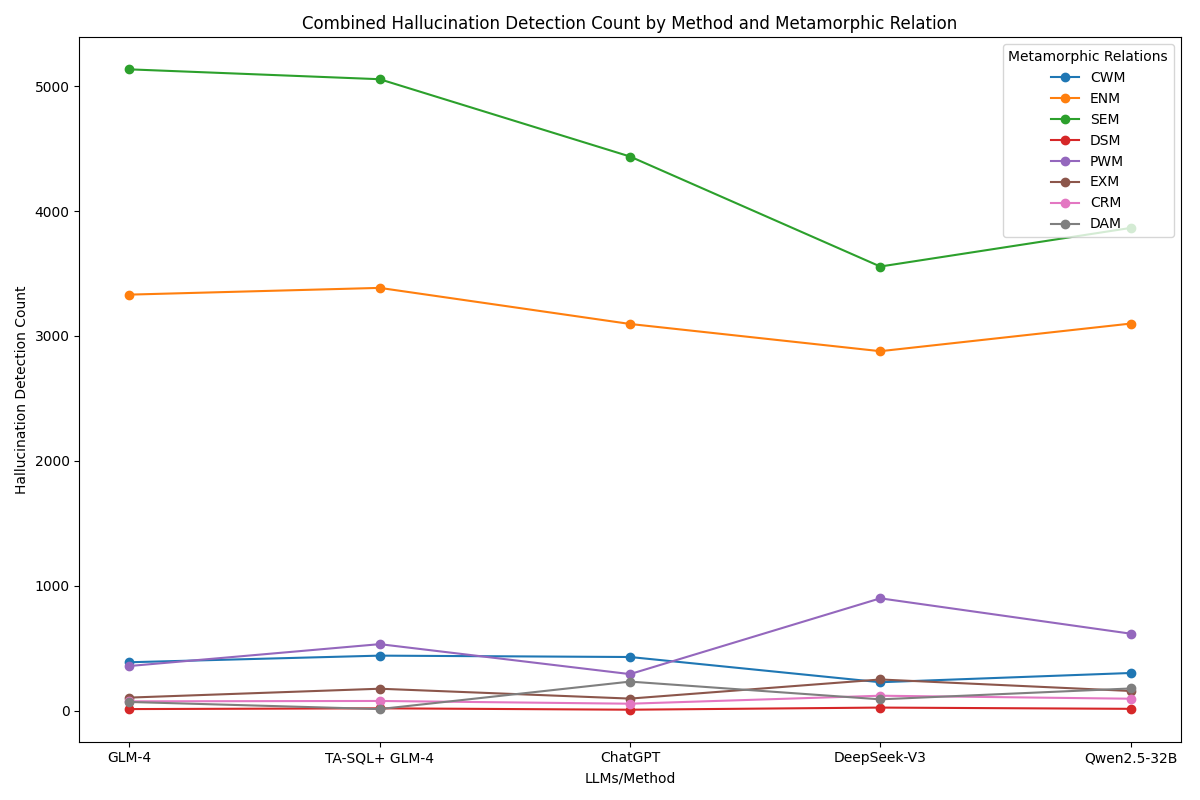}\\
	\caption{Hallucination detection by  single-class MRs} \label{fig:HDSNum}
\end{figure}

In this study, we designed eight major classes of MRs for schema-linking and logical synthesis to analyze the sensitivity of LLMs to various types of hallucinations. The number of hallucinations detected based on single-class MRs is shown in Figure~\ref{fig:HDSNum}.

As can be seen from Figure~\ref{fig:HDSNum}, the schema-linking phase is crucial in Text-to-SQL tasks as it involves mapping natural language queries to the corresponding database schema components (tables, columns, and relationships). This phase is where many hallucinations can occur due to incorrect mappings or misunderstandings of the query intent. The MRs CWM , ENM, and SEM are particularly important in this phase.

ENM and SEM: These categories show the highest number of hallucinations detected, indicating that the correct mapping of entities and schema components is a significant challenge for LLMs. For instance, in the SGLMD dataset, GLM-4 detected 594 ENM and 1019 SEM hallucinations, highlighting the importance of accurate entity and schema mapping.

DSM: This category shows a relatively low number of hallucinations, suggesting that LLMs perform well in understanding database schema changes. For example, in the SGLMD dataset, GLM-4 detected only 0 DSM hallucinations, indicating robust performance in this area.

The logical synthesis phase involves generating the final SQL query from the parsed components. This phase is where logical errors and hallucinations can occur due to incorrect JOIN operations, SQL clause application, and computational reasoning. The MRs PWM, EXM, CRM, and DAM are particularly important in this phase.

PWM and EXM: These categories show a high number of hallucinations detected, indicating that the correct mapping of predicates and expressions is challenging for LLMs. For example, in the SGLMD dataset, GLM-4 detected 63 PWM and 16 EXM hallucinations, suggesting that these areas require further improvement.

DAM: This category shows an increasing trend in the number of hallucinations detected, indicating that even though LLMs can perceive database schema changes, they still struggle with logical synthesis. For example, in the SGLMD dataset, GLM-4 detected 15 DAM hallucinations, highlighting the need for better logical reasoning in the final SQL generation.

The ENM category detected a significant number of hallucinations across all datasets, indicating that complex entity-name mapping is a major weakness of LLMs. These models struggle to accurately map entities from natural language to database schemas, which directly impacts the accuracy of the generated SQL queries. Similarly, the SEM category also detected a high number of hallucinations, further emphasizing the importance of the connection between entities and schemas. The models face challenges in understanding the relationship between entities in natural language queries and database schemas, necessitating improvements in entity recognition and semantic comprehension.

In contrast, the DSM category detected fewer hallucinations, suggesting that LLMs perform well in understanding changes in database schemas. This indicates that the models can adapt effectively to variations in database schemas, resulting in more accurate SQL query generation. The PWM category, however, detected a higher number of hallucinations, indicating that LLMs are prone to errors when processing predicate word mapping. This may be due to inaccurate judgments of sentence structure by the models, leading to the generation of more hallucinations in the output text.

The EXM category also detected a high number of hallucinations, revealing difficulties for LLMs in handling expression mapping. This could be attributed to the models' insufficient semantic understanding, which results in hallucinations when generating SQL queries. The CRM category detected a moderate number of hallucinations, indicating that LLMs encounter some difficulties in mapping column relationships. This may stem from inaccurate understanding of the relationships between columns, causing hallucinations in the generated SQL queries.

Lastly, the DAM category detected a considerable number of hallucinations, showing that LLMs still cannot completely avoid generating hallucinations during the logical synthesis stage. This may be due to logical inconsistencies in the models when processing data and parameter mapping, leading to erroneous information in the generated SQL queries.

\textbf{From the comparison results, we can observe that for all four datasets, our single-class MRs-based approach achieved higher performance. Specifically, the high number of hallucinations detected in ENM and SEM categories during the schema-linking phase indicates that accurate entity and schema mapping is crucial for reducing hallucinations. Additionally, the high number of hallucinations detected in PWM and EXM categories during the logical synthesis phase suggests that logical reasoning and expression mapping are significant areas for improvement. Our approach effectively identifies these challenges, providing valuable insights for enhancing the accuracy and reliability of Text-to-SQL models.}

\subsection{Threats to validity}
Our study is subject to two primary validity threats. Firstly, the threat to internal validity stems from our re-implementation of baseline methods. Although we adhered as closely as possible to the procedures described in the original work, inevitable discrepancies in implementation details—including preprocessing steps, hyperparameter choices, or subtle model behaviors, may introduce deviations in performance. Such differences could affect the fidelity of our replication and, consequently, the reliability of comparative results. Secondly, the threat to conclusion validity arises from the constrained empirical scope of our investigation. We evaluate SQLHD exclusively within the context of hallucination detection for Text-to-SQL generation. While this setting provides a controlled environment for assessing the method, it limits the extent to which our findings can be generalized. To robustly establish the broader applicability of SQLHD, future work should examine its effectiveness across additional code-generation or program-synthesis tasks, particularly those involving other programming languages where hallucination phenomena manifest differently.

\section{Discussion}
\label{sec:discussion}
\subsection{Comparative Effectiveness of MRs}
 From the experimental results, it is evident that certain MRs are more effective in detecting hallucinations compared to others. Specifically, in the schema-linking phase, MRs such as ESR, OS, and SC demonstrated higher effectiveness due to their ability to capture a broader range of potential inconsistencies between the source and follow-up questions. These MRs are particularly effective because they target common areas where hallucinations occur, such as misinterpretation of entities and incorrect sentence structuring, which are significant sources of errors in Text-to-SQL tasks. Conversely, MRs like ENSR and DSR showed less effectiveness. The reason for their lesser effectiveness may be attributed to their reliance on more specific and less frequent types of transformations. These MRs might not cover as wide a spectrum of potential hallucinations, leading to fewer detections. To improve the effectiveness of these MRs, it is suggested to expand their scope to include more varied and subtle types of transformations that could lead to inconsistencies.
 
 In the logical synthesis phase, MRs like PI and AROE were notably effective. This effectiveness can be attributed to their impact on the logical structure and extremum handling in SQL queries, which are critical for accurate query generation. However, MRs like CWR showed less effectiveness, indicating a need for enhanced logical reasoning capabilities in SQL generation. To address this, it is recommended to refine these MRs to better simulate real-world variations in database schema and to improve the model's adaptability to such changes.

 The experimental data highlights that certain MRs excel at identifying hallucinations. In schema linking, ESR, OS, and SC are particularly effective due to their broad detection of inconsistencies, targeting common hallucination sources like entity misinterpretation and sentence structure errors. ENSR and DSR are less effective, likely due to their focus on less common transformations, leading to fewer detections. To enhance these MRs, it's advised to broaden their scope to cover a wider array of transformations. In logical synthesis, PI, AROE are notably potent, impacting SQL's logical structure and extremum handling, crucial for accurate query generation. However, CWR is less effective, signaling a need to bolster SQL generation's logical reasoning. Refining these MRs to better reflect real-world database variations is recommended.

% \textbf{\textit{From the comparison results, we can observe that for all four datasets, our approach based on single MRs achieved higher performance in hallucination detection . This suggests that our approach provides valuable insights for improving the accuracy and reliability of Text-to-SQL models by addressing key challenges in entity recognition, semantic comprehension, and logical reasoning.}} 

\subsection{Cost and Scalability}
 Our prototype activates MRs per natural-language query, which expands to about twelve language-model calls after deduplication and batch-padding. In a cloud setting the round trip takes under three seconds and costs a few cents per question. For sporadic use this is attractive, yet nightly regression suites that process tens of thousands of questions would quickly accumulate a noticeable bill.
 
Moving the workload to a single local GPU removes the per-query charge. A twenty-billion-parameter model in four-bit precision occupies roughly ten gigabytes for weights and six for activations, so a consumer RTX 3090 with twenty-four gigabytes can hold the network and still leave room for concurrent batches. Latency remains near the cloud figure while throughput climbs to well over one thousand questions an hour. The only ongoing expense is electricity; at peak draw the card adds about one kilowatt-hour every three thousand queries, an amount that is easily budgeted in most organisations.

To stretch the throughput further we let the pipeline stop early when the first few relations already agree that no hallucination is present. This simple heuristic cuts the average number of calls in half and drops the F1-score by less than two percent, pushing the effective rate beyond two thousand questions an hour on the same card and making large-scale continuous validation economically practical.

\section{Conclusion}
\label{sec:conclusion}
This paper introduces SQLHD, a novel method for detecting hallucinations in LLMs performing Text-to-SQL tasks, leveraging MT. SQLHD splits the Text-to-SQL process into two phases: schema linking and logical synthesis, conducting hallucination detection in each to promptly identify and rectify issues, thereby markedly boosting SQL generation accuracy. Our contributions are threefold: Firstly, we propose a two-stage hallucination detection approach. Secondly, we define seventeen MRs to tackle logical and grammatical errors. Thirdly, through extensive experiments on the Spider and BIRD datasets, we demonstrate SQLHD's superior performance over existing methods. Future work will focus on developing more advanced MRs to further refine hallucination detection and incorporating real-time feedback to enhance the robustness and adaptability of LLMs in dynamic settings.

% \begin{acks}
% \end{acks}

\bibliographystyle{ACM-Reference-Format}
\bibliography{arxiv}
\end{document}